\documentclass[fleqn,twoside]{article}
\usepackage{espcrc2}

\usepackage{epsfig}
\usepackage{graphicx}

\newif\ifpreprint
\preprinttrue

\title{
\ifpreprint
\rightline{\normalsize FERMILAB-Conf-02/192-E}
\rightline{\normalsize UASLP-IF-02-008}
\vspace{0.4cm}
Two RICH Detectors as Velocity Spectrometers in the CKM
Experiment\thanks{Contributed talk at the Fourth Workshop on
RICH Detectors, June 5-10, 2002, Pylos, Greece. To be published in NIMA.}
\else
Two RICH Detectors as Velocity Spectrometers in the CKM Experiment
\fi
}
\author{
J.~Engelfried\address[slp]{Instituto de F\'{\i}sica, Universidad Aut\'onoma
de San Luis Potos\'{\i}, Mexico}\thanks{Corresponding author; 
email: {\tt jurgen@ifisica.uaslp.mx}},
P.S.~Cooper\address[fermi]{Fermi National Accelerator Laboratory, 
Batavia, IL, USA},
A.~Morelos\addressmark[slp], 
I.~Torres\addressmark[slp],
\ifpreprint
A.~Barker\address[uc]{University of Colorado, Bolder, CO},
L.~Bellantoni\addressmark[fermi],
V.~Bolotov\address[inr]{Institute of Nuclear Research, Troisk, Russia}, 
G.~Britvich\address[ihep]{Institute of High Energy Physics,
Serpukhov, Russia},
M.~Campbell\address[umich]{University of Michigan, Ann Arbor, Michigan 48109},
R.~Coleman\addressmark[fermi],
C.~Dukes\address[uv]{University of Virgina, Charlottesville, VA 22901},
J.~Frank\address[brook]{Brookhaven National Laboratory, Upton, NY, USA},
R.~Gustafson\addressmark[umich],
H.~Huang\addressmark[uc],
A.V.~Inyakin\addressmark[ihep],
C.M.~Jenkins\address[usa]{University of South Alabama, Mobile, Alabama 36688},
S.H.~Kettell\addressmark[brook],
T.R.~Kobilarcik\addressmark[fermi],
V.~Kurshetsov\addressmark[ihep],
A.~Kushnirenko\addressmark[ihep],
L.G.~Landsberg\addressmark[ihep],
K.~Lang\address[austin]{University of Texas at Austin, Austin, Texas 78712},
S.~Laptev\addressmark[inr], 
M.~Longo\addressmark[umich],
L.~Lu\addressmark[uv],
V.~Molchanov\addressmark[ihep],
K.~Nelson\addressmark[uv],
H.~Nguyen\addressmark[fermi],
R.~Niclases\addressmark[uc],
V.~Obraztsov\addressmark[ihep],
H.~Park\addressmark[umich],
A.~Pastsiak\addressmark[inr], 
S.I.~Petrenko\addressmark[ihep],
A.~Polarush\addressmark[inr], 
V.~Polyakov\addressmark[ihep],
E.~Ramberg\addressmark[fermi],
R.~Strand\addressmark[brook],
V.I.~Rykalin\addressmark[ihep],
R.~Sirodeev\addressmark[inr],
A.~Soldatov\addressmark[ihep], 
M.M.~Shapkin\addressmark[ihep],
O.G.~Tchikilev\addressmark[ihep],
R.S.~Tschirhart\addressmark[fermi],
H.B.~White\addressmark[fermi],
D.~Vavilov\addressmark[ihep],
M.~Wilking\addressmark[uc],
J.Y.~Wu\addressmark[fermi],
O.~Yushchenko\addressmark[ihep]
\else
for the CKM Collaboration
\fi
}

\begin{document}

\begin{abstract}
We present the design of two velocity spectrometers, to be used
in the recently approved CKM experiment. 
CKM's main goal is the measurement of the
branching ratio of $K^+ \to \pi^+ \nu \bar\nu$ with a precision of 
$10\,\mbox{\%}$, via decays in flight of the $K^+$.  
The design of both RICH detectors is based on the SELEX Phototube RICH. 
We will discuss the design and the expected performance, based on
studies with SELEX data and Monte Carlo Simulations.
\end{abstract}

\maketitle


\section{The Physics}
\begin{figure*}
\begin{center}
\leavevmode
\includegraphics[width=0.8\textwidth,height=8.2cm]{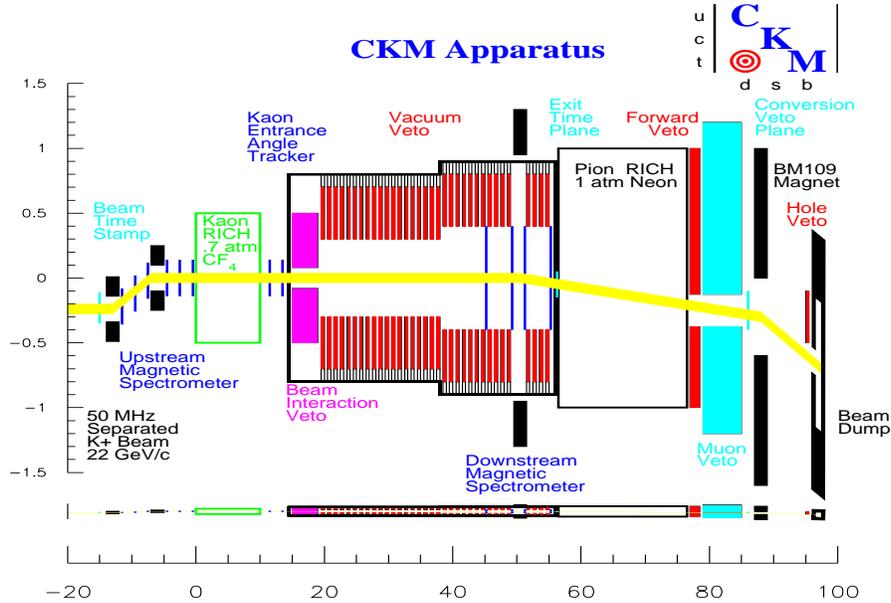}
\caption{Proposed layout of the CKM experiment. Bottom: real
scale. Top: Enlarged vertical scale.}
\label{layout}
\end{center}
\end{figure*}
The basic goal of the CKM 
\ifpreprint
experiment~\cite{propa,propb} 
\else
experiment~\cite{ckm,propa,propb} 
\fi
is to determine $|V_{td}|$, contributing to test the Standard Model
hypothesis that a single phase in the CKM matrix is the sole
source of ${\cal CP}$ violation.  The idea is to over-constrain 
the predictions with measurements.  To really falsify the hypothesis,
one needs a set of observables which can not only be measured sufficiently
well, but also need to have controlled errors in the theoretical connection
between the observable and the CKM matrix elements.
A short list of these clean measurements to be
made in the near future includes:
$B_d^0\to \Psi K_s$ (the golden mode at the $B$-Factories),
the ratio of the mixing parameters of $B_d$ and $B_s$ (to be measured
at the Tevatron),
$K^+\to\pi^+\nu\bar\nu$ and $K^0\to\pi^0\nu\bar\nu$.

The standard model prediction for the branching ratio of
$K^+\to\pi^+\nu\bar\nu$ is~\cite{Buras,ambrosio}
$0.55  \times 10^{-10}[(1.35-\bar\rho)^2+(1.05\bar\eta)^2]$,
defining an ellipse in the $\bar\rho-\bar\eta$ plane.  
The calculation~\cite{inami}
is done as a ratio to the branching ration of the decay 
$K^+\to\pi^0 e^+ \nu$, via weak isospin rotation.
With current values  this leads to $(0.72\pm0.21)\times 10^{-10}$.
The theoretical uncertainty is estimated to be around
$8\,\mbox{\%}$~\cite{Buras,ambrosio}, and dominated by the uncertainty in the
charm quark mass.  We therefore aim at observing 100~events of that decay,
giving a measurement of $|V_{td}|$ of $10\,\mbox{\%}$ including 
theoretical errors.

Recently the Brookhaven experiment~787 observed~\cite{brookhaven} two events 
of this decay after several years of running;
their branching ratio measurement is high but consistent with the
prediction of the Standard Model. A continuation of that stopped kaon
experiment, E949, hopes to see 5-10~events in the next couple of years.

\section{The Experiment}
\ifpreprint
The CKM (Charged Kaons at the Main Injector) Collaboration
\else
The CKM (Charged Kaons at the Main Injector)~\cite{ckm} Collaboration
\fi
plans to measure the branching ratio of 
$K^+\to\pi^+\nu\bar\nu$ to a statistical precision of $10\,\mbox{\%}$ by
observing the decay {\em in flight}.
After a lengthy process~\cite{loi,propa,propb} the experiment obtained
Stage~1 (physics) approval at Fermilab in June~2001.  The experiment plans
to run in parallel to the Collider Run~2.

Protons from the Fermilab Main Injector are used to produce an RF-separated
beam, containing $30\cdot10^6$ $K^+$ per second, with a total flux
of $<50\cdot10^6/\mbox{s}$. 
A layout of the experiment is shown in fig.~\ref{layout}.
The largest part of the experiment is a $50\,\mbox{m}$ 
long vacuum decay volume, filled with photon counters to veto against
(under others)
the decay $K^+\to\pi^+\pi^0$, one of the main backgrounds to suppress.
This system should give a $\pi^0$ rejection of $<1.6\cdot10^{-7}$, requiring
an additional kinematic rejection of $<3\cdot10^{-5}$ by redundant
spectrometer systems: two conventional magnetic spectrometers and two 
velocity spectrometers based on RICH detectors.  The kinematic rejection
uses the fact that the main background modes  are 
two-body decays, and, if calculating
the invariant mass the $\pi^+$ recoils to ($M_{\rm miss}$), 
results within resolution a fixed value, whereas the signal mode is a
three-body decay, leading to a distribution in $M_{\rm miss}$.
The spectrometer systems have to measure 
the magnitude of the momenta of the 
$K^+$ and $\pi^+$, as well as 
the decay angle between the $K^+$ and $\pi^+$.

\section{RICHes as Velocity Spectrometers}
The velocity spectrometers consist of RICH detectors, which will
measure the vector velocity (magnitude via the ring radius, track angles
via the position of the ring center) for the incoming $K^+$ and for the
outgoing $\pi^+$.  The detectors are modeled after the
SELEX RICH~\cite{elba,bignim,israel,pylos},
following the founding principle of the experiment
that only proven detector technologies should be used to avoid surprises
while searching for very rare decays. The results from the SELEX RICH are being
used to study resolution effects and to tune the Monte Carlo simulation.
\begin{figure}
\begin{center}
\leavevmode
\epsfxsize=\hsize
\epsffile{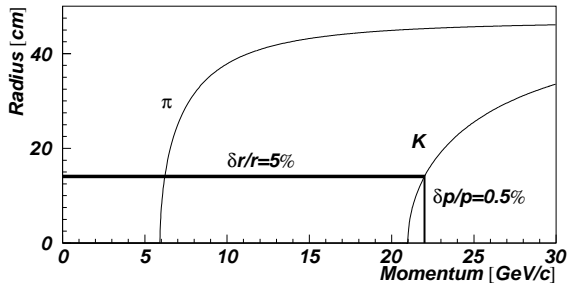}
\caption{Ring Radius as a function of momentum in the Kaon RICH}
\label{radmom}
\end{center}
\end{figure}
In fig.~\ref{radmom} we show, when operating the RICH close to threshold, that
one can obtain with a modest accuracy in ring radius an excellent measurement
of momentum. In addition, the curves for kaon and pion are widely separated,
so particle identification comes nearly for free. That a RICH can be operated
in this regime was shown by the SELEX RICH~\cite{pylos}.

The experiment requires momentum and angular resolutions in the order of
$1\,\mbox{\%}$, which can easily be translated into required ring radius
resolutions. An even more important question is how gaussian is this
resolution, e.g.\ how will non-gaussian tails in the resolution function
affect the overall performance when searching for rare decays.
\begin{figure}
\begin{center}
\leavevmode
\epsfxsize=\hsize
\epsffile{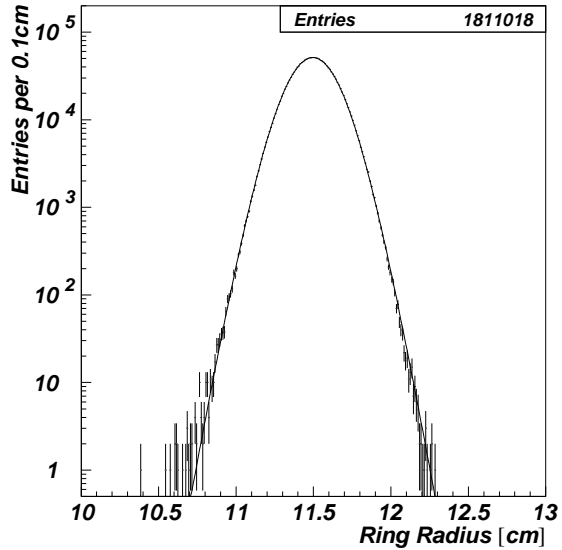}
\caption{Ring radius distribution for $\beta=1$ particles in the SELEX RICH,
using the standard SELEX algorithm.
The curve fitted is explained in the text.}
\label{06kpmg}
\end{center}
\end{figure}
In fig.~\ref{06kpmg} we show the result of a study with SELEX data, single
track events with $\beta=1$.
To assign hits to a ring, we used the standard SELEX algorithm, which uses
the information of the tracking system to determine the ring center.
The average number of hits on a ring is $13.4$,
with a Poisson distribution. We fit a sum of gaussians to this distribution,
where the width of the gaussians is scaled by $\sqrt{N-3}$ ($N$ being the
number of hits on a ring), and the weight is given by how many events we
observed for every given $N$.  As can be seen, the resolution function
shows a gaussian behavior over more than 4 orders of magnitude.

\begin{figure}
\begin{center}
\leavevmode
\epsfxsize=\hsize
\epsffile{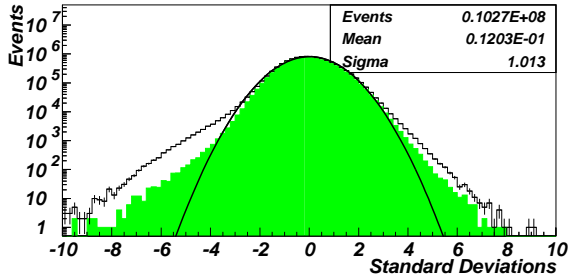}
\caption{Pull distribution for the ring radius for $\beta=1$ particles in 
the SELEX RICH. A simple standalone pattern recognition is used.  The two
histograms show no and a soft cut.}
\label{fig63}
\end{center}
\end{figure}

For CKM we do not want to use information from the tracking system
in order to keep
the two spectrometer systems really independent and redundant. We used 
the same SELEX sample as above, but this time performing a ring fit
assigning {\em all} observed hits to the ring. We calculated the
pull (deviation from mean divided by error) for the radius,
where we scaled again the
error with $\sqrt{N-3}$. The result is shown in fig.~\ref{fig63}.
The gaussian behavior found in this study is already sufficient for
the experiment.  We are currently working on improving the pattern recognition
algorithm. In addition, the integration in SELEX was $170\,\mbox{nsec}$,
and in CKM we expect a time resolution in the order of a few~nsec, which should
help to reduce the noise contribution.

\section{The Pion RICH}
The Pion RICH will be a straight copy of the SELEX RICH, with the only
exception of doubling the vessel length from $10\,\mbox{m}$ to
$20\,\mbox{m}$, with a corresponding change in curvature of the mirrors
to $40\,\mbox{m}$. The radiator gas will be Neon at atmospheric pressure.
Monte Carlo simulations showed that we should use 
$1/2\,\mbox{in.\ }$photomultipliers sensitive down to $160\,\mbox{nm}$ to
obtain a
momentum resolution
in the $1\,\mbox{\%}$ range. The resolution is balanced between the
different contributions
(pm size, chromatic dispersion), and
the $\pi$-$\mu$ separation is $>10\,\sigma$.

\section{The Kaon RICH}
\begin{figure*}
\begin{center}
\leavevmode
\epsfxsize=12cm
\epsffile{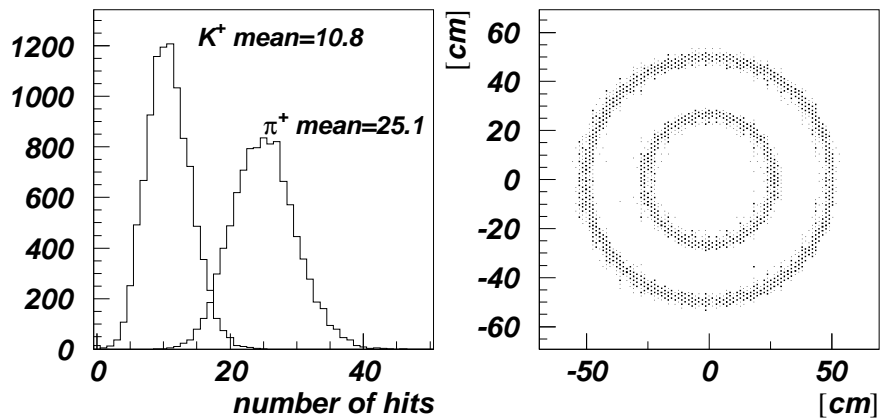}
\caption{Results of Monte Carlo simulations for the Kaon RICH:
Left: Number of photons detected for beam kaons and pions. Right: 
Distribution of detected photons in the focal plane for 
10000 beam $K^+$ and 10000 beam $\pi^+$ events.}
\label{kaonrich}
\end{center}
\end{figure*}

Due to the beam characteristics (about $10\,\mbox{cm}$ diameter, very small
divergence) 
only a small number of phototubes is needed to cover the focal plane of the
Kaon RICH.
Due to space restrictions and rate problems, we will double-fold the
light path in a $11\,\mbox{m}$ long vessel, filled either with N$_2$
at atmospheric pressure or with CF$_4$ at about $0.7\,\mbox{atm}$. 
The spherical mirror
($40\,\mbox{m}$ radius) and the second flat mirror will be outside the
beam.
The mirror sizes are chosen so that only part of the
Cherenkov photons for the (bigger) pion ring are reflected, but the number
of detected photons is still enough, as shown in fig.~\ref{kaonrich}.
The rings from kaons and pions are nicely separated, a fact we will
use in the experiment trigger. The rates per tube will be as high
as $1.4\,\mbox{MHz}$.  The resolution is limited by chromatic dispersion,
and we will use $1/2\,\mbox{in.\ }$photomultipliers with a cutoff
around $300\,\mbox{nm}$.

\section{Resolution Studies with MC}
We implemented both RICHes, together with all other detectors,
into GEANT~\cite{geant}.  We ran studies with up to $10^6$ $K^+\to\pi^+\pi^0$
events with full kinematic reconstruction to verify the
performance of single detectors
and of the experiment as a whole.   Detailed results can be found
in~\cite{propb}.  A main result is that we have no correlations
between the two spectrometer systems (magnetic, RICHes).
With the current design we should be able to reach the
goal of the experiment.

\section{Summary}
The recently approved CKM experiment will measure the branching ratio
of $K^+\to\pi^+\nu\bar\nu$ with 100~events, to obtain a $10\mbox{\%}$
measurement, including theoretical uncertainties, of the magnitude of
$V_{td}$.  We will employ two phototube RICH detectors, modeled after the
SELEX RICH, as velocity spectrometers.  We can stand high rates, have 
a very low noise rate, good time resolution (all intrinsic properties
of photomultipliers), and we expect the non-gaussian tails of the
resolution functions to be sufficiently small.

\section*{Acknowledgments}
This work was supported in part by
Consejo Nacional de Ciencia y Tecnolog\'{\i}a (Mexico),
FAI-UASLP,
and the US Department of Energy 
\ifpreprint
(contract No.\ DE-AC02-76CHO3000 and DE-AC02-98CH10886).
\else
(contract No.\ DE-AC02-76CHO3000).
\fi

\end{document}